\documentclass{article}
\usepackage{amsfonts}
\usepackage{makeidx}
\usepackage{latexsym,amsmath,amssymb,amscd}
\usepackage{makecell}
\usepackage{xcolor}

\allowdisplaybreaks
\newtheorem{theorem}{Theorem}

\begin{document}
\title{Integrable and superintegrable potentials of 2d autonomous
conservative dynamical systems}
\author{Antonios Mitsopoulos$^1$\thanks{Email: antmits@phys.uoa.gr}, Michael Tsamparlis$^1$\thanks{%
Email: mtsampa@phys.uoa.gr} and Andronikos Paliathanasis$^{2,3}$\thanks{%
Email: anpaliat@phys.uoa.gr} \\
{\ \ }\\
{\textit{$^1$Faculty of Physics, Department of
Astronomy-Astrophysics-Mechanics,}}\\
{\ \textit{University of Athens, Panepistemiopolis, Athens 157 83, Greece.}}%
\\
{\textit{$^2$Institute of Systems Science, Durban University of Technology}}
\\
{\textit{Durban 4000, Republic of South Africa.}}
\\
{\textit{$^3$Instituto de Ciencias F\'{\i}sicas y Matem\'{a}ticas,}} \\
{\textit{Universidad Austral de Chile, Valdivia, 5090000, Chile.}} }
\maketitle

\begin{abstract}
We consider the generic quadratic first integral (QFI) of the form $%
I=K_{ab}(t,q)\dot{q}^{a}\dot{q}^{b}+K_{a}(t,q)\dot{q}^{a}+K(t,q)$ and
require the condition $dI/dt=0.$ The latter results in a system of partial
differential equations which involve the tensors $K_{ab}(t,q)$, $K_{a}(t,q)$%
, $K(t,q)$ and the dynamical quantities of the dynamical equations.
These equations divide in two sets. One set
which involves only geometric quantities of the configuration space and a second set
which contains the interaction of these quantities with the dynamical fields.
 A theorem is presented which provides a systematic solution
of the system of equations in terms of the collineations of the kinetic
metric in the configuration space. This solution being geometric and covariant, applies
to higher dimensions and curved spaces. The results are applied to the simple
but interesting case
of two-dimensional (2d) autonomous conservative Newtonian potentials. It is found that there
are two classes of 2d integrable potentials and that superintegrable
potentials exist in both classes. We recover most main results, which have
been obtained by various methods, in a single and
systematic way.

\end{abstract}

\section{Introduction}

\label{Introduction}

The precise meaning of the solution of a system of differential equations
can be cast in several ways \cite{anleach}. We say that we have
determined a closed-form solution for a dynamical system when we have
determined a set of explicit functions describing the variation of the
dependent variables in terms of the independent variable(s). On the other hand,
when we have proved the existence of a sufficient number of independent
explicit first integrals and invariants for the dynamical system, we
say that we have found an analytic solution of the dynamical equations.
In addition, an algebraic solution is found when one has proved the
existence of a sufficient number of explicit transformations which permit
the reduction of the system of differential equations to a system of
algebraic equations. A feature, central to each of these three equivalent
prescriptions of integrability, is the existence of explicit functions which are
 first integrals/invariants or the coefficient functions of
the aforementioned transformations.

A first integral (FI) of a dynamical system is a scalar $I$ defined on the
phase space of the system such that $dI/dt=0$. The FIs are classified
according to the power of the momenta. The linear FIs (LFIs) are linear in
the momenta, the quadratic FIs (QFIs) contain products of two momenta and so
on. A dynamical system of $n$ degrees of freedom is called integrable if it
admits $n$ (functionally) independent FIs which are in involution \cite%
{Arnold 1989}, that is, their Poisson brackets are zero, i.e. $%
\{I_{i},I_{j}\}=0$. The maximum number of independent FIs that a dynamical
system of $n$ degrees of freedom can have is $2n-1$ and when this is the
case an integrable system is called superintegrable. The above apply to all
dynamical systems which are described by dynamical equations independently
if they are Lagrangian, or Hamiltonian. If the dynamical system is
Hamiltonian, then the FIs are defined equivalently by the requirement $\{H,I\}=0$ where $H$ is the Hamiltonian function of the system.

FIs are important for the determination of the solution and the study of dynamical systems.
In particular, when a dynamical system is integrable, then (in principle) the
solution of the dynamical equations can be found by means of quadratures.
Such dynamical systems are characterized as Liouville integrable. For this
reason the systematic computation of FIs is a topic of active interest for a
long time, perhaps by the time of the early Mechanics. Originally the FIs
were concerned in the field of the geometry of surfaces (see for example
\cite{Darboux 1901}) where an attempt was made to compute all integrable and
superintegrable 2d surfaces. A new dimension to the topic gave the
introduction of the theorem of Noether in 1918 \cite{Noether 1918} which
prevails the topic since then. More recently one more systematic method, but
less general than the Noether one, was presented in which one assumes a
general form of the QFI and then uses the condition $dI/dt=0$, or the $%
\{H,I\}=0$, to find a system of simultaneous equations involving the
coefficients defining $I$ (see for instance \cite{Katzin 1973}, \cite{Katzin 1974}, \cite{Kalotas}, \cite{Fris}). The solution of that system of conditions provides us with all the QFIs admitted by a given dynamical system.

The determination of integrable and superintegrable systems is a topic which
is in continuous investigation. Obviously a universal method which computes
the FIs for all types of dynamical equations independently of their
complexity and degrees of freedom is not available. For this reason the
existing studies restrict their considerations to flat spaces or spaces of
constant curvature of low dimension (e.g. \cite{Darboux 1901}, \cite%
{Whittaker 1959}, \cite{Dorizzi Grammatikos Ramani 1983}, \cite{Thompson
1984a}, \cite{Sen}, \cite{TsaPal1}, \cite{Ranada 1997}, \cite{Kalnins 2001}
and references therein). The prevailing cases involve the autonomous
conservative dynamical systems with two degrees of freedom and the
classification of the potential functions in integrable and superintegrable.
A comprehensive review of the known integrable and superintegrable 2d
autonomous potentials is given in \cite{Hietarinta 1987}.

Besides the two methods mentioned above, other approaches have appeared. For
example Koenigs \cite{Koenigs} used coordinate transformations in order to
solve the system of equations resulting from the condition $\{H,I\}=0$. This
solution of that system of equations gives the general functional form of
the QFIs and the superintegrable free Hamiltonians, that is the ones which
possess two more QFIs - in addition to the Hamiltonian - which are
functionally independent. Koenig's method has been generalized in several
works (see \cite{Daskalogiannis 2006} and references cited therein) for two
dimensional autonomous conservative systems.

In the present work we follow the method which uses the solution of the
simultaneous system of equations resulting from the condition $dI/dt=0.$
This approach has been used extensively, e.g. \cite{Katzin 1973}, \cite{Katzin 1974}, \cite{Kalotas}, however always for special cases only. In this
work we use Theorem \ref{The first integrals of an autonomous holonomic dynamical
system} proved in \cite{Tsamp2020} which gives the general solution of this system in terms of the
collineations and the Killing tensors (KTs) of the kinetic metric in the configuration
space. This solution is systematic and covariant therefore can be used in
higher dimensions and for curved configuration spaces. Furthermore it is shown that
it is directly related to the Noether approach.

Theorem \ref{The first integrals of an autonomous holonomic dynamical
system} is applied to the case of 2d autonomous conservative dynamical
systems in order to determine the integrable and the superintegrable
potentials. It is found that the integrable potentials are classified in
\textbf{Class I} and \textbf{Class II} and that superintegrable potentials
exist in both classes. All potentials together with their QFIs are listed in
tables for easy reference. All the results listed in the review paper of
\cite{Hietarinta 1987} as well as in more recent works (e.g. \cite{Ranada
1997}, \cite{Kalnins 2001}) are recovered while some
new ones are found which admit time-dependent LFIs and QFIs.

\section{Gauged Noether symmetries and QFIs}

\label{sec.genconsider}

We consider an autonomous conservative dynamical system of $n$ degrees of
freedom $q^{a}$ with kinetic energy $T=\frac{1}{2}\gamma _{ab}\dot{q}^{a}%
\dot{q}^{b}$ where $\dot{q}^{a}=\frac{dq^{a}}{dt}$. We define in the
configuration space of the system the kinetic metric $\gamma _{ab}$ by the
requirement $\gamma _{ab}=\frac{\partial ^{2}T}{\partial \dot{q}^{a}\dot{q}%
^{b}}.$ When the dynamical system is regular, that is, $\det \left( \frac{%
\partial ^{2}T}{\partial \dot{q}^{a}\dot{q}^{b}}\right) \neq 0,$ it can be
shown that the dynamical equations can be written in the form%
\begin{equation}
\ddot{q}^{a}=-\Gamma _{bc}^{a}\dot{q}^{b}\dot{q}^{c}-V(q)^{,a}
\label{eq.Noe0}
\end{equation}%
where $\Gamma_{bc}^{a}$ are the Riemann connection coefficients defined by the
kinetic metric $\gamma_{ab}$, $V(q)$ stands for the conservative forces, a comma indicates the partial derivative and the Einstein summation convention is used. Finally the metric $\gamma_{ab}$ is used for lowering and raising the indices.

The main methods for the determination of the FIs of (\ref{eq.Noe0}) are: \newline
a. The theorem of Noether which is the standard one and requires a
Lagrangian. \newline
b. The direct method (see e.g. \cite{Katzin 1973}, \cite{Katzin 1974}, \cite%
{Kalotas}, \cite{Daskalogiannis 2006}, \cite{Karpathopoulos 2018}) which is
not applied widely, uses only the dynamical equations and involves the
solution of the system of equations resulting from the condition $dI/dt=0$.

The two methods are related as follows.

In the Noether approach the Noether symmetries are generated by vector fields%
\footnote{%
We restrict our considerations to vector fields in the jet space $J^{1}(t,q,%
\dot{q}).$}
\begin{equation*}
\mathbf{X}=\xi (t,q,\dot{q})\frac{\partial }{\partial t}+\eta ^{a}(t,q,\dot{q%
})\frac{\partial }{\partial q^{a}}
\end{equation*}%
whose first prolongation $\mathbf{X}^{[1]}$ in the jet space $J^{1}(t,q,\dot{%
q})$ is given by
\begin{equation}
\mathbf{X}^{[1]}=\xi \partial _{t}+\eta ^{a}\partial _{q^{a}}+\left( \dot{%
\eta}^{a}-\dot{q}^{a}\dot{\xi}\right) \partial _{\dot{q}^{a}}.
\label{FL.10.3}
\end{equation}

Let $L=T-V$ be the Lagrangian of the dynamical system. The Noether
symmetries of (\ref{eq.Noe0}) are its Lie symmetries which satisfy in
addition the Noether condition
\begin{equation}
\mathbf{X}^{\left[ 1\right] }L+L\dot{\xi}=\dot{f}
\label{GenHolonNoether'sFinCond}
\end{equation}%
where $f(t,q,\dot{q})$ is the Noether or the gauge function. According to
the theorem of Noether a Noether symmetry produces the FI
\begin{equation}
I=f-L\xi -\frac{\partial L}{\partial \dot{q}^{a}}\left( \eta ^{a}-\xi \dot{q}%
^{a}\right) .  \label{GenHolonNoether's1Integr}
\end{equation}

The Noether symmetries are classified in a formal way in two classes\footnote{%
The original paper of Noether does not distinguish these classes. For a
recent enlightening discussion of Noether theorem see \cite{Hadler Paliathanasis Leach 2018} and references therein.}:\newline
a) The point Noether symmetries whose generators are vector fields on the
augmented configuration space $\{t,q^{a}\}$ and usually lead to LFIs.
\newline
b) The generalized Noether symmetries whose generators are vector fields in
the jet space $J^{1}(t,q,\dot{q})$ which produce FIs of higher degree.%
\newline
In the present work we restrict our considerations to generalized Noether
symmetries in the first jet space $J^{1}(t,q,\dot{q})$ which produce QFIs.

The 2d autonomous potentials which admit point Noether symmetries have
already been classified in \cite{Sen} and more recently recovered and
extended in \cite{TsaPal1}. Furthermore in \cite{TsaPal1} it has been shown
that the generators of the point Noether symmetries are the elements of the
homothetic algebra of the kinetic metric. Obviously that firm result is not
expected to apply in the case of generalized Noether symmetries which form
an infinite dimensional Lie group.

It is well-known \cite{Stephani book ODES} that the generalized Noether
symmetries have one extra degree of freedom (being special generalized Lie symmetries)
which is removed if we consider the gauge condition $\xi =0$, which we
assume to be the case. Therefore, the (gauged) Noether symmetries we
consider are generated by vector fields of the form $X=\eta ^{a}(q,\dot{q},%
\ddot{q},..)\frac{\partial }{\partial q^{a}}$ and accordingly the Noether condition and
the corresponding FI are simplified as follows
\begin{equation}
X^{[1]}L=\dot{f},\quad I=f-\frac{\partial L}{\partial \dot{q}^{a}}\eta ^{a}.
\label{GenHolonNoether's2Cond2}
\end{equation}

In the direct method one assumes for the QFI the generic expression
\begin{equation}
I=K_{ab}(t,q)\dot{q}^{a}\dot{q}^{b}+K_{a}(t,q)\dot{q}^{a}+K(t,q)
\label{eq.Noe1}
\end{equation}%
where $K_{ab}(t,q)$, $K_{a}(t,q)$, $K(t,q)$ are unknown tensor quantities
and demands the condition\footnote{{Equivalently, if the system is
Hamiltonian, one requires $\{H,I\}-\frac{\partial I}{\partial t}=0$ where $\{.,.\}$ is the Poisson bracket.}} $\frac{dI}{dt}=0.$ This condition leads to a
system of simultaneous equations among the coefficients $K_{ab}(t,q)$, $%
K_{a}(t,q)$, $K(t,q)$ whose solution provides all the LFIs and the QFIs of
the system of this functional form. The involvement of the specific dynamical system is in the
replacement of the term $\ddot{q}^{a}$ whenever it appears from the dynamical equations (\ref%
{eq.Noe0}).

The direct approach is related to the Noether symmetries because once one
has determined the QFI the generator of the corresponding gauged Noether
symmetry and the Noether function follow immediately. Indeed for a gauged
Noether symmetry (in the gauge $\xi =0$) relation (\ref%
{GenHolonNoether's1Integr}) becomes%
\begin{equation}
I=f-\frac{\partial L}{\partial \dot{q}^{a}}\eta ^{a}.  \label{eq.Noe2}
\end{equation}%
Replacing $L=T-V(q)$ we find%
\begin{equation}
I=f-\eta ^{a}\gamma _{ab}\dot{q}^{b}=f-\eta _{a}\dot{q}^{a}  \label{eq.Noe3}
\end{equation}%
and using (\ref{eq.Noe1}) it follows%
\begin{equation}
\eta _{a}=-K_{ab}\dot{q}^{b}-K_{a},\enskip f=K  \label{eq.Noe4}
\end{equation}%
that is we obtain directly the Noether generator and the Noether function
from the QFI $I$ by reading the coefficients $K_{ab}(t,q)$, $K_{a}(t,q)$ and
$K(t,q)$ respectively. It can be proved that: (a) the set $\{-K_{ab}\dot{q}%
^{b}-K_{a};K\}$ does satisfy the gauged Noether condition $\mathbf{X}^{[1]}L=\dot{f}$ and (b) the QFI $I$ defined in (\ref{eq.Noe1}) is not in
general Noether invariant (as it is the case for the point Noether symmetries - see proposition 2.2 in \cite{Sarlet 1981}). Finally, the (gauged) point Noether symmetries which are defined
by the vector $K_{a}$ ($K_{ab}=0)$ give the LFIs whereas the (gauged) generalized Noether symmetries with $K_{ab}\neq 0$ give the QFIs.

\section{The QFIs of an autonomous conservative dynamical system}
\label{Theorem}

It is known (see e.g. \cite{Katzin 1973}, \cite{Kalotas}) that condition $dI/dt=0$ leads to the following system of equations\footnote{
Using the dynamical equations (\ref{eq.Noe0}) to replace $\ddot{q}^{a}$ whenever it appears the condition $dI/dt=0$ is written
\[
K_{(ab;c)}\dot{q}^{a}\dot{q}^{b}\dot{q}^{c}+\left(
K_{ab,t}+K_{(a;b)}\right) \dot{q}^{a}\dot{q}^{b}+\left(
K_{a,t}+K_{,a}-2K_{ab}V^{,b} \right) \dot{q}^{a} +K_{,t}-K_{a}V^{,a}=0.
\]
}
\begin{eqnarray}
K_{(ab;c)} &=&0  \label{eq.veldep4.1} \\
K_{ab,t}+K_{(a;b)} &=&0  \label{eq.veldep4.2} \\
-2K_{ab}V^{,b}+K_{a,t}+K_{,a} &=&0  \label{eq.veldep4.3} \\
K_{,t}-K_{a}V^{,a} &=&0.  \label{eq.veldep4.4}
\end{eqnarray}
Here round/square brackets indicate symmetrization/antisymmetrization of the enclosed indices and a semi-colon denotes the Riemannian covariant derivative. In the special case of a scalar function, for example the potential $V$, it holds that $V^{,a}=V^{;a}$.

Condition $K_{(ab;c)}=0$ implies that $K_{ab}$ is a Killing tensor (KT) of
order 2 (possibly zero) of the kinetic metric $\gamma _{ab}$. Because $%
\gamma _{ab}$ is autonomous we assume
\begin{equation*}
K_{ab}(t,q)=g(t)C_{ab}(q)
\end{equation*}%
where $g(t)$ is an arbitrary analytic function and $C_{ab}(q)$ ($C_{ab}=C_{ba}$) is a KT of order 2 of the metric $\gamma _{ab}.$ This choice
of $K_{ab}$ and equation (\ref{eq.veldep4.2}) indicate that we set
\begin{equation*}
K_{a}(t,q)=f(t)L_{a}(q)+B_{a}(q)
\end{equation*}%
where $f(t)$ is an arbitrary analytic function and $L_{a}(q)$, $B_{a}(q)$
are arbitrary vectors. With these choices the system of equations (\ref%
{eq.veldep4.1}) -(\ref{eq.veldep4.4}) becomes
\begin{eqnarray}
g(t)C_{(ab;c)} &=&0  \label{eq.veldep5} \\
g_{,t}C_{ab}+f(t)L_{(a;b)}+B_{(a;b)} &=&0  \label{eq.veldep6} \\
-2g(t)C_{ab}V^{,b}+f_{,t}L_{a}+K_{,a} &=&0  \label{eq.veldep7} \\
K_{,t}-(fL_{a}+B_{a})V^{,a} &=&0.  \label{eq.veldep8}
\end{eqnarray}

Conditions (\ref{eq.veldep5}) - (\ref{eq.veldep8}) must be supplemented with
the integrability conditions $K_{,at}=K_{,ta}$ and $K_{,[ab]}=0$ for the scalar $K.$ The integrability condition $K_{,at}=K_{,ta}$ gives -
if we make use of (\ref{eq.veldep7}) and (\ref{eq.veldep8}) - the PDE
\begin{equation}
f_{,tt}L_{a}+f_{,t}L_{b}A_{a}^{b}+f\left( L_{b}V^{,b}\right) _{;a}+\left(
B_{b}V^{,b}\right) _{;a}-2g_{,t}C_{ab}V^{,b}=0.  \label{eq.veldep9}
\end{equation}

Condition $K_{,[ab]}=0$ gives the equation known as the second order
Bertrand-Darboux PDE
\begin{equation}
2g\left( C_{[a\left\vert c\right\vert }V^{,c}\right) _{;b]}-f_{,t}L_{\left[
a;b\right] }=0  \label{eq.veldep10}
\end{equation}
where indices enclosed between vertical lines are overlooked by symmetrization or antisymmetrization symbols.

Finally, the system of equations which we have to solve consists of
equations (\ref{eq.veldep5}) - (\ref{eq.veldep10}). The general solution of
that system in terms of the collineations of the kinetic metric is given in
the following Theorem (see  \cite{Tsamp2020}).

\begin{theorem}
\label{The first integrals of an autonomous holonomic dynamical system}

The functions $g(t),f(t)$ are assumed to be analytic so that they may be
represented by polynomial expansion as follows
\begin{equation}  \label{eq.thm1}
g(t) = \sum^n_{k=0} c_k t^k = c_0 + c_1 t + ... + c_n t^n
\end{equation}
\begin{equation}  \label{eq.thm2}
f(t) = \sum^m_{k=0} d_k t^k = d_0 + d_1 t + ... + d_m t^m
\end{equation}
where $n, m \in \mathbb{N}$, or may be infinite, and $c_k, d_k \in \mathbb{R}
$. Then the independent QFIs of an autonomous conservative dynamical system
are the following: \bigskip

\textbf{Integral 1.}
\begin{equation*}
I_{1} = -\frac{t^{2}}{2} L_{(a;b)}\dot{q}^{a}\dot{q}^{b} + C_{ab}\dot{q}^{a} \dot{q}^{b} + t L_{a} \dot{q}^{a} + \frac{t^{2}}{2} L_{a}V^{,a} + G(q)
\end{equation*}
where $C_{ab}$, $L_{(a;b)}$ are KTs, $\left(L_{b}V^{,b}\right)_{,a} =
-2L_{(a;b)} V^{,b}$ and $G_{,a}= 2C_{ab}V^{,b} - L_{a}$.

\textbf{Integral 2.}
\begin{equation*}
I_{2} = -\frac{t^{3}}{3} L_{(a;b)}\dot{q}^{a}\dot{q}^{b} + t^{2} L_{a} \dot{q%
}^{a} + \frac{t^{3}}{3} L_{a}V^{,a} - t B_{(a;b)} \dot{q}^{a}\dot{q}^{b} +
B_{a}\dot{q}^{a} + tB_{a}V^{,a}
\end{equation*}
where $L_{a}$, $B_{a}$ are such that $L_{(a;b)}$, $B_{(a;b)}$ are KTs, $%
\left(L_{b}V^{,b}\right)_{,a} = -2L_{(a;b)} V^{,b}$ and $\left(B_{b}V^{,b}%
\right)_{,a} = -2B_{(a;b)} V^{,b} - 2L_{a}$.

\textbf{Integral 3.}
\begin{equation*}
I_{3} = -e^{\lambda t} L_{(a;b)}\dot{q}^{a}\dot{q}^{b} + \lambda e^{\lambda t} L_{a} \dot{q}^{a} + e^{\lambda t} L_{a} V^{,a}
\end{equation*}
where $\lambda \neq 0$, $L_{a}$ is such that $L_{(a;b)}$ is a KT and $\left(L_{b}V^{,b}
\right)_{,a} = -2L_{(a;b)} V^{,b} - \lambda^{2} L_{a}$.
\end{theorem}

It can be checked that the FIs listed above produce all the potentials which
admit a LFI or a QFI given in \cite{TsaPal1} and are due to point Noether symmetries.
Since, as shown above,  these FIs also follow form a gauged velocity dependent Noether symmetry we conclude  that \emph{there does not exist a one-to-one correspondence between Noether FIs and the type of Noether symmetry.} For example the FI of the total energy (Hamiltonian) $E= \frac{1}{2} \gamma_{ab}\dot{q}^{a}\dot{q}^{b} +V(q)$ (case \textbf{Integral 1} for $L_{a}=0$ and $C_{ab}=\frac{\gamma_{ab}}{2}$) is generated by the point Noether symmetry $\Big(\xi=1, \eta_{a}=0; f=0\Big)$ and also by the gauged generalized Noether symmetry $\Big(\xi=0, \eta_{a}= -\frac{1}{2}\gamma_{ab}\dot{q}^{b}; f=V(q) \Big)$.

The FI $-I_{2}(L_{a}=0)$ for $B_{a}$ be a HV with conformal factor $\psi=const$ is generated by the point Noether symmetry $\Big( \xi=2\psi t, \eta_{a}=B_{a}; f=ct \Big)$ such that $B_{a}V^{,a} +2\psi V +c =0$; and also by the gauged generalized Noether symmetry $\Big( \xi=0, \eta_{a}= -t\psi\gamma_{ab}\dot{q}^{b} +B_{a}; f=-tB_{a}V^{,a} \Big)$.

As a final example we consider the FI $-\frac{I_{3}}{\lambda}$ for the gradient HV $L_{a}=\Phi(q)_{,a}$ where  $\Phi_{;ab}= \psi\gamma_{ab}$ with $\psi=const$. This FI is generated by the point Noether symmetry
\[
\Big( \xi= \frac{2\psi}{\lambda}e^{\lambda t}, \eta_{a}= e^{\lambda t} \Phi(q)_{,a}; f= \lambda e^{\lambda t} \Phi(q) -\frac{c}{\lambda} e^{\lambda t} \Big)
\]
where $\lambda, c$ are non-zero constants and $\Phi_{,a}V^{,a}= -2\psi V -\lambda^{2}\Phi +c$; and also by the gauged generalized Noether symmetry
\[
\Big( \xi=0, \eta_{a} = - \frac{e^{\lambda t} }{\lambda} \psi\gamma_{ab}\dot{q}^{b} + e^{\lambda t} \Phi_{,a}; f= - \frac{e^{\lambda t}}{\lambda} \Phi_{,a} V^{,a} \Big).
\]

\section{The determination of the QFIs}

\label{The determination of the QFIs}

From Theorem \ref{The first integrals of an autonomous holonomic dynamical
system} follows that for the determination of the QFIs the following
problems have to be solved:

a. Determine the KTs of order 2 of the kinetic metric $\gamma_{ab}$.

b. Determine the special subspace of KTs of order 2 of the form $%
C_{ab}=L_{(a;b)}$ where  $L^{a}$ is a vector.

c. Determine the KTs satisfying the constraint $G_{,a}= 2C_{ab} V^{,b}$.

d. Find all KVs $L_{a}$ of the kinetic metric which satisfy the constraint $%
L_{a}V^{,a}=s$ where $s$ is a constant, possibly zero.

We note that constraints a. and b. depend only on the kinetic metric.
Because the kinetic energy is a positive definite non-singular quadratic
2-form we can always choose coordinates in which this form reduces either to
$\delta _{ab}$ or to $A(q)\delta _{ab}.$ Since we know the KTs and all the
collineations of a conformally flat metric (of Euclidean or Lorentzian
character) \cite{Rani 2003} we already have the results for all
autonomous (Newtonian or special relativistic) conservative dynamical systems.

The involvement of the potential function is only in the constraints c.
and d. which also depend on the geometric characteristics of the kinetic
metric. There are two different ways to proceed.

\subsection{The potential $V\left(q\right)$ is known}

\label{subsec.pot.given}

In this case the following procedure is used.\newline

a) Substitute $V$ in the constraints $L_{a}V^{,a}=s$ and $%
G_{,a}=2C_{ab}V^{,b}$ and find conditions for the defining parameters of $L_{a}$ and $C_{ab}$. \newline

b) From these conditions determine $L_{a}$, $C_{ab}$. \newline

c) Substitute $C_{ab}$ in the constraint $G_{,a}=2C_{ab}V^{,b}$ and
find the function $G(q)$. \newline

d) Using the above results write the FI $I$ in each case and determine
directly the gauged Noether generator and the Noether function. \newline

e) Examine if $I$ can be reduced to simpler independent FIs or if it is
new. \newline

\subsection{The potential $V\left(q\right)$ is unknown}

\label{subsec.pot.not.given}

In this case the following algorithm is used.\newline

a) Compute the KTs and the KVs of the kinetic metric. \newline

b) Solve the PDE $L_{a}V^{,a}=s$ or the\footnote{
The integrability conditions for the scalar $G$ are very general PDEs
from which one can find only special solutions by making additional
simplifying assumptions (e.g. symmetries) involving $L_{a}$, $C_{ab}$ and $V(q)$ itself. Therefore one does not find the most general solution. For example in \cite{Markakis 2014} it is required that the QFI\ $I\ $is
axisymmetric, that is $\phi ^{\lbrack 1]}I=0$ where $\phi
^{i[1]}=-y\partial_{x}+x\partial_{y} -\dot{y}\partial_{\dot{x}} +\dot{x}%
\partial_{\dot{y}}$ is the first prolongation of the rotation $\phi
^{i}=-y\partial x+x\partial y$. It is proved easily that in this case we
have also the constraints $L_{\phi}K_{a}=0$ and $L_{\phi}K_{ab}=0$.}
 $G_{,[ab]}=0$ and find the
possible potentials $V(q)$. \newline

c) Substitute the potentials and the KTs found in the constraint $G_{,a}=2C_{ab}V^{,b}$ and compute the function $G(q)$. \newline

d) Write the FI $I$ for each potential and determine the gauged Noether
generator and the Noether function. \newline

e) Examine if $I$ can be reduced further to simpler independent FIs or if it is a new FI. \newline

In the following sections we assume the potential is not given and apply the
second procedure. For that we need first the geometric quantities of the 2d Euclidean plane $E^2$.

\section{The geometric quantities of $E^{2}$}

\label{sec.E2.geometry}

Using well-known results (see also \cite{Thompson 1984a}, \cite%
{Karpathopoulos 2018}) \ we state the following:

- $E^{2}$ admits two gradient Killing vectors (KVs) $\partial _{x},\partial _{y}$ whose
generating functions are $x,y$ respectively and one non-gradient KV (the
rotation) $y\partial _{x}-x\partial _{y}$. These vectors can be written
collectively%
\begin{equation}
L_{a}=\left(
\begin{array}{c}
b_{1}+b_{3}y \\
b_{2}-b_{3}x%
\end{array}%
\right)  \label{FL.15}
\end{equation}%
where $b_{1},b_{2},b_{3}$ are arbitrary constants, possibly zero.

- The general KT of order 2 in $E^{2}$ is
\begin{equation}
C_{ab}=\left(
\begin{array}{cc}
\gamma y^{2}+2\alpha y+A & -\gamma xy-\alpha x-\beta y+C \\
-\gamma xy-\alpha x-\beta y+C & \gamma x^{2}+2\beta x+B%
\end{array}%
\right)  \label{FL.14b}
\end{equation}%
from which follows%
\begin{equation}
C_{ab}(q)\dot{q}^{a}\dot{q}^{b}=\left( \gamma y^{2}+2\alpha y+A\right) \dot{x}%
^{2}+2\left( -\gamma xy-\alpha x-\beta y+C\right) \dot{x}\dot{y}+\left( \gamma
x^{2}+2\beta x+B\right) \dot{y}^{2}  \label{FL.10.2}
\end{equation}
where $\alpha, \beta, \gamma, A, B, C$ are arbitrary constants.

- The vectors $L^{a}$ generating KTs of $E^{2}$ of the form $%
C_{ab}=L_{(a;b)}$ are
\begin{equation}
L^{a}=\left(
\begin{array}{c}
-2\beta y^{2}+2\alpha xy+Ax+a_{1}y+a_{4} \\
-2\alpha x^{2}+2\beta xy+a_{3}x+By+a_{2}%
\end{array}%
\right)  \label{FL.14}
\end{equation}
where $a_{1}, a_{2}, a_{3}, a_{4}$ are  arbitrary constants.

- The KTs $C_{ab}=L_{(a;b)}$ in $E^{2}$ generated from the vector (\ref%
{FL.14}) are%
\begin{equation}
C_{ab}=L_{(a;b)}=\left(
\begin{array}{cc}
L_{x,x} & \frac{1}{2}(L_{x,y}+L_{y,x}) \\
\frac{1}{2}(L_{x,y}+L_{y,x}) & L_{y,y}%
\end{array}%
\right) =\left(
\begin{array}{cc}
2\alpha y+A & -\alpha x-\beta y+C \\
-\alpha x-\beta y+C & 2\beta x+B%
\end{array}%
\right)  \label{FL.14.1}
\end{equation}%
where\footnote{%
Note that $L^{a}$ in (\ref{FL.14}) is the sum of the non-proper ACs of $%
E^{2} $ and not of its KVs which give $C_{ab}=0.$ .} $2C=a_{1}+a_{3}$.
Observe that these KTs are special cases of the general KTs (\ref{FL.14b}%
) for $\gamma =0$.

According to Theorem \ref{The first integrals of an autonomous holonomic
dynamical system} the above are common to all 2d Newtonian systems and what
changes in each particular case are the constraints $G_{,a}=2C_{ab}V^{,b}$
and $L_{a}V^{,a}=s$ which determine the potential $V(q)$.

\section{Computing the potentials and the FIs}

\label{sec.find.Pots}

The application of Theorem \ref{The first integrals of an autonomous
holonomic dynamical system} in the case of $E^{2}$ indicates that there are
three different ways to find potentials that admit QFIs (other than the
Hamiltonian): \bigskip

1) The constraint $L_{a}V^{,a}=s$ which leads to the PDE
\begin{equation}
(b_{1}+b_{3}y)V_{,x}+(b_{2}-b_{3}x)V_{,y}-s=0.  \label{eq.PDE1}
\end{equation}

2) The constraint $G_{,a}=2C_{ab}V^{,b}$ which leads to the second order
Bertrand-Darboux PDE ($G_{,xy}=G_{,yx}$)
\begin{eqnarray}
0 &=&(\gamma xy+ \alpha x+\beta y-C)(V_{,xx}-V_{,yy})+\left[ \gamma
(y^{2}-x^{2})-2\beta x+2\alpha y+A-B\right] V_{,xy}-  \notag \\
&&-3(\gamma x+\beta )V_{,y}+3(\gamma y+\alpha)V_{,x}.  \label{eq.PDE2}
\end{eqnarray}

3) The constraint $\left( L_{b}V^{,b}\right) _{,a}=-2L_{(a;b)}V^{,b}-\lambda
^{2}L_{a}$ with\footnote{%
For $\lambda=0$ this constraint is a subcase of $G_{,a}=2C_{ab}V^{,b}$ hence
only the case $\lambda \neq 0$ must be considered.} $\lambda \neq 0$ and the integrability
condition $\left( L_{b}V^{,b}\right) _{,xy}=\left( L_{b}V^{,b}\right) _{,yx}$
which lead to the PDEs%
\begin{eqnarray}
0 &=&(-2\beta y^{2}+2\alpha xy+ Ax+ a_{1}y+a_{4})V_{,xx}+(-2\alpha x^{2}+2\beta
xy+ a_{3}x+By+a_{2})V_{,xy}+  \notag \\
&&+(-6\alpha x+2a_{3}+a_{1})V_{,y}+3(2\alpha y+A)V_{,x}+\lambda ^{2}(-2\beta y^{2}+2\alpha xy+Ax +a_{1}y+a_{4})  \label{eq.PDE3.1}
\end{eqnarray}
\begin{eqnarray}
0 &=&(-2\alpha x^{2}+2\beta xy +a_{3}x +By +a_{2})V_{,yy}+(-2\beta y^{2}+2\alpha xy+Ax +a_{1}y +a_{4})V_{,xy}+  \notag \\
&&+3(2\beta x+B)V_{,y}+(-6\beta y+2a_{1}+a_{3})V_{,x}+\lambda
^{2}(-2\alpha x^{2} +2\beta xy +a_{3}x+By+a_{2})  \label{eq.PDE3.2}
\end{eqnarray}%
\begin{eqnarray}
0 &=&(\alpha x+\beta y-C)(V_{,xx}-V_{,yy})+\left( -2\beta x +2\alpha y+A-B\right)V_{,xy} -3\beta V_{,y} +3\alpha V_{,x}+  \notag \\
&&+\frac{\lambda ^{2}}{2}(6\alpha x-6\beta y+a_{1}-a_{3}), \quad 2C=a_{1}+a_{3}.
\label{eq.PDE3.3}
\end{eqnarray}

For $\alpha=\beta =0$ and $a_{1}=a_{3}$ equation (\ref{eq.PDE3.3}) reduces to (%
\ref{eq.PDE2}). Therefore in order to find new potentials one of these
conditions must be relaxed. This case of finding potentials is the most
difficult because the problem is over-determined, i.e. we have a system of
three PDEs (\ref{eq.PDE3.1})-(\ref{eq.PDE3.3}) and only one unknown function, the
$V(x,y)$.

In the following sections we solve these constraints and find the admitted
potentials which, as a rule, are integrable. Subsequently we apply Theorem %
\ref{The first integrals of an autonomous holonomic dynamical system} to
each of these potentials in order to compute the admitted FIs and determine
which of those are integrable and in particular superintegrable.

\section{The constraint $L_{a}V^{,a} = s$}

\label{sec.const1}

The constraint $L_{a}V^{,a}=s$ gives (\ref{eq.PDE1}) which can be solved
using the method of the characteristic equation.

To cover all possible occurrences we have to consider the following cases:
a) $b_{3}=0$ and $b_{1}\neq 0$ (KVs $\partial _{x}$ and $\partial
_{x},\partial _{y})$; b) $b_{3}=b_{1}=0$ and $b_{2}\neq 0$ (KV $\partial
_{y})$; and c) $b_{3}\neq 0$ ( KVs $y\partial _{x}-x\partial _{y}$; $%
\partial _{x},y\partial _{x}-x\partial _{y}$; and $\partial _{y},y\partial
_{x}-x\partial _{y})$. For each case the solution is shown in the following
table: \bigskip

\begin{tabular}{|c|c|c|}
\hline
Case & KV & $V(x,y)$ \\ \hline
a & $b_{3}=0,b_{1}\neq 0$ & $\frac{s}{b_{1}}x+F(b_{1}y-b_{2}x)$ \\
b & $b_{3}=b_{1}=0$, $b_{2}\neq 0$ & $\frac{s}{b_{2}}y+F(x)$ \\
c & $b_{3}\neq 0$ & $\frac{s}{b_{3}}\tan ^{-1}\left( \frac{y+\frac{b_{1}}{%
b_{3}}}{-x+\frac{b_{2}}{b_{3}}}\right) +F(b_{1}y+\frac{b_{3}}{2}y^{2}-b_{2}x+%
\frac{b_{3}}{2}x^{2})$ \\ \hline
\end{tabular}
\bigskip

We shall refer to the above solutions as \textbf{Class I} potentials. In order
to determine if these potentials admit QFIs we apply Theorem \ref{The first
integrals of an autonomous holonomic dynamical system} to the following
potentials resulting from the table above:
\begin{eqnarray*}
V_{1} &=&cx+F(y-bx) \\
V_{2} &=&cy+F(x)
\end{eqnarray*}%
\begin{equation*}
V_{3}=c\tan ^{-1}\left( \frac{y+b_{1}}{-x+b_{2}}\right) +F\left( \frac{%
x^{2}+y^{2}}{2}+b_{1}y-b_{2}x\right) .
\end{equation*}%
Before we continue we recall that \emph{if $I_{1},I_{2},...,I_{k}$ are FIs of a
given dynamical system then any function $f(I_{1},...,I_{k})$ is also a FI of the dynamical system.}

\subsection{The potential $V_{1}=cx+F(y-bx)$}

\label{subsec.V1}

\textbf{Case a.} $b=0$ and $F=\lambda y$.

The potential reduces to $V_{1a}=cx + \lambda y$.

The irreducible FIs are%
\begin{equation*}
L_{11}=\dot{x}+ct,\enskip L_{12}=\dot{y}+\lambda t,\enskip Q_{11}=\frac{1}{2}%
\dot{x}^{2}+cx,\enskip Q_{12}=\frac{1}{2}\dot{y}^{2}+\lambda y.
\end{equation*}%
We note that $Q_{11}+Q_{12}=\frac{1}{2}(\dot{x}^{2}+\dot{y}^{2})+V=H$ the
Hamiltonian. We compute $\{Q_{11},Q_{12}\}=0$, $\{L_{11},Q_{11}\}=-c$.

The FIs $I_{1}=Q_{11}+Q_{12}$, $I_{2}=\lambda L_{11}-cL_{12}=\lambda \dot{x}%
-c\dot{y} $ and $I_{3}=Q_{11}$ are functionally independent and satisfy the
relations
\begin{equation*}
\{I_{1},I_{2}\}=\{I_{1},I_{3}\}=0,\enskip\{I_{2},I_{3}\}= -c\lambda.
\end{equation*}
Therefore the potential $V_{1a}$ is superintegrable.

We note that the FIs $I_{2}$, $Q_{11}$ are respectively the FIs (3.1.4) and
(3.2.20) of \cite{Hietarinta 1987}. \bigskip

\textbf{Case b.} $\frac{d^{2}F}{dw^{2}}\neq 0$ and $w\equiv y-bx$.

The irreducible FIs are
\begin{equation*}
L_{21}=\dot{x}+b\dot{y}+ct,\enskip L_{22}=t(\dot{x}+b\dot{y})-(x+by)+\frac{c%
}{2}t^{2},\enskip Q_{21}=(\dot{x}+b\dot{y})^{2}+2c(x+by).
\end{equation*}

For $F(y-bx)=-\frac{1}{2}\lambda ^{2}y^{2}$ and $b=0$ we have the potential $%
V_{1b}=cx-\frac{1}{2}\lambda ^{2}y^{2}$, $\lambda \neq 0$, which admits the
additional time-dependent FI $L_{23}=e^{\lambda t}(\dot{y}-\lambda y)$.
Observe also that in this case $Q_{21}$ reduces to $Q_{e1}=\frac{1}{2}\dot{x}%
^{2}+cx$ which using the Hamiltonian generates the QFI
\begin{equation*}
Q_{e2}\equiv H-Q_{e1}=\frac{1}{2}\dot{y}^{2}-\frac{1}{2}\lambda ^{2}y^{2}.
\end{equation*}

The LFI $L_{21}(c=0)$ is the (3.1.4) of \cite{Hietarinta 1987}.

We compute $\{H,L_{21}\}=\frac{\partial L_{21}}{\partial t}=c$ because $%
L_{21}$ is a time-dependent FI.

The potential of the case b is integrable because $\{H,Q_{21}\}=0$.

Moreover
\begin{equation*}
\{H,L_{22}\} = L_{21} = \frac{\partial L_{22}}{\partial t}, \quad
\{L_{21},L_{22}\}= 1 + b^{2},
\end{equation*}
\begin{equation*}
\{Q_{21},L_{21}\} = 2c(1+b^{2}) = 2c\{L_{21},L_{22}\}, \quad
\{Q_{21},L_{22}\}= 2(1+b^{2})L_{21}=2\{L_{21},L_{22}\}L_{21}.
\end{equation*}

For the special case $V=cx-\frac{1}{2}\lambda ^{2}y^{2}$ we have
\begin{equation*}
\{H,L_{23}\}=\{Q_{e2},L_{23}\}=\lambda L_{23}=\frac{\partial L_{23}}{%
\partial t},\enskip \{Q_{e1},L_{23}\}=0.
\end{equation*}%
The triplet $Q_{e1},Q_{e2},L_{23}$ proves that this potential is
superintegrable.\newline
We note that in \cite{Hietarinta 1987} only the \textbf{Class II} potentials
(to be considered in the next section) are examined for superintegrability
(see \cite{Hietarinta 1987} p.108 (3.2.34)-(3.2.36) ).

\bigskip

For $c\neq 0$ the potential $V_{1}=cx+F(y-bx)$ is not included in \cite%
{Hietarinta 1987} because the author seeks for autonomous LFIs of the
form (3.1.1) and in that case $s=0$.

\subsection{The potential $V_{2}=cy+F(x)$}

\label{subsec.V2}

We consider the case $F^{\prime \prime }= \frac{d^{2}F}{dx^{2}}\neq 0$ because otherwise we
retrieve the potential $V_{1a}$ discussed above.

The irreducible FIs are%
\begin{equation*}
L_{31}=\dot{y}+ct,\enskip Q_{31}=\frac{1}{2}\dot{x}^{2}+F(x),\enskip Q_{32}=%
\frac{1}{2}\dot{y}^{2}+cy.
\end{equation*}%
Therefore the potential $V_{2}$ is integrable. This potential is also of the
form $V=F_{1}(x)+F_{2}(y)$, which is the (3.2.20) of \cite{Hietarinta 1987}.

For $F(x)=- \frac{1}{2} \lambda^{2}x^{2}$ we obtain the potential $V_{2a}=cy-%
\frac{1}{2}\lambda^{2}x^{2}$, $\lambda\neq0$, which admits the additional FI
$L_{32}= e^{\lambda t}(\dot{x}-\lambda x)$. This potential is
superintegrable because of the functionally independent triplet $Q_{31}$, $%
Q_{32}$ and $L_{32}$.

\subsection{The potential $V_{3} = c \tan^{-1}\left( \frac{y+b_{1}}{-x +b_{2}%
} \right) + F\left( \frac{x^{2}+ y^{2}}{2} + b_{1}y - b_{2}x \right)$}

\label{subsec.V3}

We find the time-dependent LFI
\begin{equation*}
L_{51}=y\dot{x}-x\dot{y}+b_{1}\dot{x}+b_{2}\dot{y}+ct.
\end{equation*}
For $c=0$ this potential is integrable. For $c\neq 0$ we do not know.

- For $c=0$ and $F=\lambda \left(\frac{x^{2}+y^{2}}{2}+ b_{1}y-b_{2}x\right)$%
, $\lambda \neq 0$, the independent FIs are%
\begin{equation*}
L_{41}=y\dot{x}-x\dot{y}+b_{1}\dot{x}+b_{2}\dot{y},\enskip Q_{41}=\frac{1}{2}%
\dot{x}^{2}+\frac{1}{2}\lambda x^{2}-\lambda b_{2}x,\enskip Q_{42}=\frac{1}{2%
}\dot{y}^{2}+\frac{1}{2}\lambda y^{2}+\lambda b_{1}y,
\end{equation*}%
\begin{equation*}
Q_{43}=\dot{x}\dot{y}+\lambda (xy+b_{1}x-b_{2}y).
\end{equation*}%
Observe that $Q_{41}+Q_{42}=H$ is the energy of the system. The LFI $L_{41}$
is the (3.1.6) of \cite{Hietarinta 1987}. The functionally independent
triplet $H$, $L_{41}$, $Q_{41}$ proves that this potential is
superintegrable. We have
\begin{equation*}
\{H,L_{41}\}=\{H,Q_{41}\}=0,\enskip\{L_{41},Q_{41}\}=-Q_{43}+\lambda
b_{1}b_{2}.
\end{equation*}

If $b_{1}=b_{2}=0$ and $\lambda = -k^{2}\neq0$ we obtain the superintegrable%
\footnote{%
A subcase of the above superintegrable potential is the potential $V_{3a}=\lambda \left(
\frac{x^{2}+y^{2}}{2}+b_{1}y-b_{2}x\right)$.} potential $V_{3b} =-\frac{1}{2}%
k^{2} (x^{2}+ y^{2})$ which admits the additional time-dependent LFIs
\begin{equation*}
L_{42\pm}=e^{\pm k t} (\dot{x} \mp k x), \enskip L_{43\pm}= e^{\pm k t}(\dot{y} \mp k y).
\end{equation*}
We also compute
\begin{equation*}
\{L_{41},Q_{42}\}=Q_{43}-\lambda b_{1}b_{2},\enskip\{L_{41},Q_{43}%
\}=2Q_{41}-2Q_{42}+\lambda (b_{2}^{2}-b_{1}^{2})
\end{equation*}%
\begin{equation*}
\{Q_{41},Q_{42}\}=0,\enskip\{Q_{41},Q_{43}\}=\{Q_{43},Q_{42}\}= -\lambda
L_{41}.
\end{equation*}

\bigskip

In section 4 of \cite{Adlam 2007} the author has found the superintegrable
\textbf{Class I} potentials $V_{1a}$ and $V_{3a}$. \bigskip

We note that in the review \cite{Hietarinta 1987} the time-dependent LFIs of the
potentials $V_{1a}$, $V_{2}$ are not discussed. In general in
\cite{Hietarinta 1987} all the time-dependent FIs are ignored, although they can
be used to decide the superintegrability of the system.

\subsection{Summary}

\label{sec.class1}

We collect the results for the \textbf{Class I} potentials in the following
tables. \bigskip

\begin{tabular}{|l|l|l|}
\hline
{\large Potential} & {\large Ref \cite{Hietarinta 1987} } & {\large LFIs and
QFIs} \\ \hline
$V_{3}(c\neq 0)=c\tan ^{-1}\left( \frac{y+b_{1}}{-x+b_{2}}\right) +F\left(
\frac{x^{2}+y^{2}}{2}+b_{1}y-b_{2}x\right) $ & - & $L_{51}=y\dot{x}-x\dot{y}%
+b_{1}\dot{x}+b_{2}\dot{y}+ct$ \\ \hline
\multicolumn{3}{|c|}{\large Integrable potentials} \\ \hline
$V_{1}=cx+F(y-bx)$, $\frac{d^{2}F}{dw^{2}}\neq 0$, $w\equiv y-bx$ & - & %
\makecell[l]{$L_{21}=\dot{x}+b\dot{y}+ct$, \\
$L_{22}=t(\dot{x}+b\dot{y})-(x+by)+\frac{c}{2}t^{2}$, \\
$Q_{21}=(\dot{x}+b\dot{y})^{2}+2c(x+by)$} \\ \hline
$V_{2}=cy+F(x)$, $F^{\prime \prime }\neq 0$ & (3.2.20) & %
\makecell[l]{$L_{31}=\dot{y}+ct$, $Q_{31}=\frac{1}{2}\dot{x}^{2}+F(x)$, \\
$Q_{32}=\frac{1}{2}\dot{y}^{2}+cy$} \\ \hline
$V_{3}(c=0)$ & (3.1.6) & $L_{51}(c=0)$ \\ \hline
\end{tabular}

\bigskip

\begin{tabular}{|l|l|l|}
\hline
\multicolumn{3}{|c|}{\large Superintegrable potentials} \\ \hline
{\large Potential} & {\large Ref \cite{Hietarinta 1987} } & {\large LFIs and
QFIs} \\ \hline
$V_{1a}=cx+\lambda y$ & \makecell[l]{(3.1.4), \\ (3.2.20)} & %
\makecell[l]{$L_{11}=\dot{x}+ct$, $L_{12}= \dot{y} + \lambda t$, \\ $Q_{11}=
\frac{1}{2}\dot{x}^{2} + cx$, $Q_{12}= \frac{1}{2}\dot{y}^{2} + \lambda y$}
\\ \hline
$V_{1b}=cx-\frac{1}{2}\lambda ^{2}y^{2}$, $\lambda \neq 0$ & (3.2.20) & %
\makecell[l]{$L_{11}$, $L_{22}(b=0)=t\dot{x}-x + \frac{c}{2}t^{2}$,
$L_{23}=e^{\lambda t}(\dot{y}-\lambda y)$, \\
$Q_{2e1}=\frac{1}{2}\dot{x}^{2}+cx$,
$Q_{2e2}=\frac{1}{2}\dot{y}^{2}-\frac{1}{2}\lambda ^{2}y^{2}$} \\ \hline
$V_{2a}=cy -\frac{1}{2}\lambda ^{2}x^{2}$, $\lambda \neq 0$ & (3.2.20) & $%
L_{31}$, $Q_{31a}=\frac{1}{2}\dot{x}^{2} -\frac{1}{2}\lambda ^{2}x^{2}$, $%
Q_{32}$, $L_{32}=e^{\lambda t}(\dot{x}-\lambda x)$ \\ \hline
$V_{3a}=\lambda \left( \frac{x^{2}+y^{2}}{2}+b_{1}y-b_{2}x\right) $, $%
\lambda \neq 0$ & (3.1.6) & \makecell[l]{$L_{41}=y\dot{x}-x\dot{y}+b_{1}
\dot{x}+b_{2}\dot{y}$, $Q_{41}= \frac{1}{2}\dot{x}^{2}+ \frac{1}{2}\lambda
x^{2}-\lambda b_{2}x$, \\ $Q_{42}=\frac{1}{2}\dot{y}^{2}+\frac{1}{2}\lambda
y^{2}+\lambda b_{1}y$, $Q_{43}=\dot{x}\dot{y}+\lambda (xy+b_{1}x-b_{2}y)$}
\\ \hline
$V_{3b}=-\frac{1}{2}k^{2}(x^{2}+y^{2})$, $k\neq 0$ & (3.1.5) & %
\makecell[l]{$L_{41b}=y\dot{x}-x\dot{y}$, $Q_{41b}= \frac{1}{2}\dot{x}^{2}-
\frac{1}{2}k^{2} x^{2}$, \\ $Q_{42b}= \frac{1}{2}\dot{y}^{2} - \frac{1}{2}
k^{2} y^{2}$, $Q_{43b}=\dot{x}\dot{y} - k^{2}xy$, \\
$L_{42\pm}=e^{\pm kt}(\dot{x}\mp kx)$, $L_{43\pm}=e^{\pm kt}(\dot{y} \mp ky)$} \\ \hline
\end{tabular}

\section{The constraint $G_{,a}=2C_{ab}V^{,b}$}

\label{sec.const2}

In this case we have the PDE (\ref{eq.PDE2})
\begin{eqnarray}
0 &=&(\gamma xy+\alpha x+\beta y-C)(V_{,xx}-V_{,yy})+\left[ \gamma(y^{2}-x^{2})-2\beta x+2\alpha y+A-B\right] V_{,xy}-  \notag \\
&&-3(\gamma x+\beta)V_{,y}+3(\gamma y+\alpha)V_{,x}.  \label{eq.Hie2}
\end{eqnarray}%
The potentials which follow from this equation we call \textbf{Class II}
potentials. This equation cannot be solved in full generality (see also \cite%
{Hietarinta 1987}), therefore we consider various cases which produce the
known FIs. We emphasize that the potentials we find in this section
are only a subset of the possible potentials which will follow from the
general solution of (\ref{eq.Hie2}). However the important point here is
that we recover the known results with a direct and unified approach which
can be used in the future by other authors to discover new integrable and
superintegrable potentials in $E^{2}$ and in other spaces.
\bigskip

1) $\gamma \neq 0$, $A=B$ and $\alpha=\beta =C=0$. Then $C_{ab}=\left(
\begin{array}{cc}
\gamma y^{2}+A & -\gamma xy \\
-\gamma xy & \gamma x^{2}+A%
\end{array}%
\right) $ and equation (\ref{eq.Hie2}) becomes
\begin{equation}
xy(V_{,xx}-V_{,yy})+(y^{2}-x^{2})V_{,xy}-3xV_{,y}+3yV_{,x}=0
\label{eq.Hie3a}
\end{equation}%
whose solution gives%
\begin{equation}
V_{21}= \frac{F_{1}\left( \frac{y}{x}\right) }{d_{1}x^{2}+d_{2}y^{2}}%
+F_{2}(x^{2}+y^{2})  \label{eq.Hie3b}
\end{equation}
where $d_{1}, d_{2}$ are arbitrary constants.

- For the subcase $d_{1}=d_{2}=1$ with $A=0$ we find the QFI
\begin{equation}
I_{11}= (y\dot{x}-x\dot{y})^{2}+ 2F_{1}\left( \frac{y}{x}\right) =(r^{2}\dot{%
\theta})^{2}-\Phi (\theta )  \label{eq.Hie3bb}
\end{equation}%
where $r^{2}=x^{2}+y^{2}$ and $\theta =\tan ^{-1}\left( \frac{y}{x}\right)$.
This is the well-known \textbf{Ermakov - Lewis invariant}; see also (3.2.11)
of \cite{Hietarinta 1987}.

- For $d_{1}\neq0$ the potential (\ref{eq.Hie3b}) is written equivalently
\begin{equation*}
V_{21} = \frac{F_{1}\left(\frac{y}{x}\right)}{x^{2}+cy^{2}} +
F_{2}(x^{2}+y^{2})
\end{equation*}
where $c$ is an arbitrary constant.

This potential admits QFIs for $F_{1}= \frac{cky^{2}+ k x^{2}}{%
x^{2}+(2-c)y^{2}}$. Therefore
\begin{equation*}
V_{21a}= \frac{k}{x^{2}+(2-c)y^{2}} + F_{2}(x^{2}+y^{2})= \frac{k}{%
x^{2}+\ell y^{2}} + F_{2}(x^{2}+y^{2})
\end{equation*}
with the QFI
\begin{equation}
I_{11a} = (y\dot{x}-x\dot{y})^{2} + \frac{2k(c-1) y^{2}}{x^{2}+(2-c)y^{2}}=
(y\dot{x}-x\dot{y})^{2} + \frac{2k(1-\ell) y^{2}}{x^{2}+\ell y^{2}}
\label{eq.Hie1a}
\end{equation}
where $\ell \equiv 2-c$.

- For $d_{1}=0$, $d_{2}\neq0$ the potential $V_{21}$ becomes
\begin{equation*}
V_{21} = \frac{F_{1}\left(\frac{y}{x}\right)}{y^{2}} + F_{2}(x^{2}+y^{2}).
\end{equation*}
This potential admits QFIs for $F_{1}= \frac{ky^{2}}{2x^{2}+y^{2}}$. Then
\begin{equation*}
V_{21b} = \frac{k}{2x^{2}+y^{2}} + F(x^{2}+y^{2})
\end{equation*}
with the QFI
\begin{equation}
I_{11b} = (y\dot{x}-x\dot{y})^{2} + \frac{ky^{2}}{2x^{2}+ y^{2}}.
\label{eq.Hie1b}
\end{equation}
Observe that $V_{21b}$ is of the form $V_{21a}(c=3/2)$ or $V_{21a}(\ell=1/2)$
with $\bar{k}\equiv 2k$. Therefore $V_{21b}$ is included in case $V_{21a}$.
\bigskip

2) $\gamma =1$ and $\alpha= \beta =B=C=0$. Then $C_{ab}=\left(
\begin{array}{cc}
y^{2}+A & -xy \\
-xy & x^{2}%
\end{array}%
\right) $ and equation (\ref{eq.Hie2}) becomes
\begin{equation}
xy(V_{,xx}-V_{,yy})+(y^{2}-x^{2}+A)V_{,xy}-3xV_{,y}+3yV_{,x}=0.
\label{eq.Hie3c}
\end{equation}

- For $A=0$ equation (\ref{eq.Hie3c}) reduces to (\ref{eq.Hie3a}). \bigskip

- For $A\neq 0$ the PDE (\ref{eq.Hie3c}) gives the \textbf{Darboux solution}
\begin{equation}
V_{22}= \frac{F_{1}(u)-F_{2}(v)}{u^{2}-v^{2}}  \label{eq.Hie3d}
\end{equation}%
where $r^{2}=x^{2}+y^{2}$, $u^{2}=r^{2}+A+\left[ (r^{2}+A)^{2}-4Ax^{2}\right]
^{1/2}$ and $v^{2}=r^{2}+A-\left[ (r^{2}+A)^{2}-4Ax^{2}\right] ^{1/2}$.

We find the QFI (see (3.2.9) of \cite{Hietarinta 1987}).
\begin{equation}
I_{21}=(y\dot{x}-x\dot{y})^{2}+A\dot{x}^{2}+\frac{v^{2}F_{1}(u)-u^{2}F_{2}(v)%
}{u^{2}-v^{2}}.  \label{eq.Hie3dd}
\end{equation}%
\bigskip

3) $\gamma =1$, $B=-A$, $C=\pm iA\neq 0$ and $\alpha=\beta =0$. Then
\begin{equation*}
C_{ab}=\left(
\begin{array}{cc}
y^{2}+A & -xy\pm iA \\
-xy\pm iA & x^{2}-A%
\end{array}%
\right)
\end{equation*}%
and equation (\ref{eq.Hie2}) gives again a potential of the form (\ref%
{eq.Hie3d}), but with $u^{2}=r^{2}+\left[ r^{4}-4A(x\pm iy)^{2}\right]
^{1/2} $ and $v^{2}=r^{2}-\left[ r^{4}-4A(x\pm iy)^{2}\right] ^{1/2}$.

We find the QFI (see (3.2.13) of \cite{Hietarinta 1987})
\begin{equation}
I_{31}=(y\dot{x}-x\dot{y})^{2}+A(\dot{x}\pm i\dot{y})^{2}+\frac{%
v^{2}F_{1}(u)-u^{2}F_{2}(v)}{u^{2}-v^{2}}.  \label{eq.Hie3db}
\end{equation}%
\bigskip

4a) $\alpha=1$ and $\beta =\gamma =A=B=C=0$. Then
\begin{equation*}
C_{ab}=\left(
\begin{array}{cc}
2y & -x \\
-x & 0%
\end{array}%
\right)
\end{equation*}%
and equation (\ref{eq.Hie2}) becomes
\begin{equation}
x(V_{,xx}-V_{,yy})+2yV_{,xy}+3V_{,x}=0  \label{eq.Hie4a}
\end{equation}%
which gives the potential
\begin{equation}
V_{24}=\frac{F_{1}(r+y)+F_{2}(r-y)}{r}  \label{eq.Hie4b}
\end{equation}%
where $r^{2}=x^{2}+y^{2}$.

We find the QFI (see (3.2.15) of \cite{Hietarinta 1987})
\begin{equation}
I_{41}=\dot{x}(y\dot{x}-x\dot{y})+\frac{(r+y)F_{2}(r-y)- (r-y)F_{1}(r+y)}{r}.
\label{eq.Hie4c}
\end{equation}
\bigskip

4b) $\beta=1$ and $\alpha =\gamma =A=B=C=0$. Then
\begin{equation*}
C_{ab}=\left(
\begin{array}{cc}
0 & -y \\
-y & 2x%
\end{array}%
\right)
\end{equation*}%
and equation (\ref{eq.Hie2}) becomes
\begin{equation}
y(V_{,xx}-V_{,yy})-2xV_{,xy}-3V_{,y}=0  \label{eq.Hie4a.1}
\end{equation}%
which gives the potential
\begin{equation}
V_{24b}=\frac{F_{1}(r+x)+F_{2}(r-x)}{r}  \label{eq.Hie4b.2}
\end{equation}%
where $r^{2}=x^{2}+y^{2}$.

We find the QFI
\begin{equation}
I_{41b}=\dot{y}(x\dot{y}-y\dot{x})+\frac{(r+x)F_{2}(r-x)-(r-x) F_{1}(r+x)}{r}
\label{eq.Hie4c.3}
\end{equation}

\emph{\ Observe that the potential (\ref{eq.Hie4b.2}) is just the (\ref%
{eq.Hie4b}) after the rotation $x\leftrightarrow y$. All the results of the
case 4b can be derived from the case 4a if we apply the transformation $%
x\leftrightarrow y$. For this reason the case 4b is ignored when we search
for integrable systems, but in superintegrability the PDE (\ref{eq.Hie4a.1})
shall be proved useful (see superintegrable potential (\ref{eq.Hie12d}) in
subsection \ref{sec.super}).}
\bigskip

5) $\alpha=1$, $\beta =-i$, $A=-B=\frac{i}{4}$, $C=\frac{1}{4}$ and $\gamma =0$.
Then
\begin{equation*}
C_{ab}=\left(
\begin{array}{cc}
2y+\frac{i}{4} & -x+iy+\frac{1}{4} \\
-x+iy+\frac{1}{4} & -2ix-\frac{i}{4}%
\end{array}%
\right)
\end{equation*}%
and equation (\ref{eq.Hie2}) becomes
\begin{equation}
(x-iy-\frac{1}{4})\left( V_{,xx}-V_{,yy}\right) +2\left( y+ix+\frac{i}{4}%
\right) V_{,xy}+3iV_{,y}+3V_{,x}=0.  \label{eq.Hie5a}
\end{equation}%
This is written equivalently
\begin{equation}
(x-iy)\left( \partial _{x}+i\partial _{y}\right) ^{2}V-\frac{1}{4}(\partial
_{x}-i\partial _{y})^{2}V+3(\partial _{x}+i\partial _{y})V=0
\label{eq.Hie5aa}
\end{equation}%
and gives the potential
\begin{equation}
V_{25} =w^{-1/2}\left[ F_{1}(z+\sqrt{w})+F_{2}(z-\sqrt{w})\right]
\label{eq.Hie5b}
\end{equation}%
where $z=x+iy$ and $w=x-iy$.

We find the QFI (see (3.2.17) of \cite{Hietarinta 1987})
\begin{eqnarray*}
I_{51} &=&(y\dot{x}-x\dot{y})(\dot{x}+i\dot{y})+\frac{i}{8}(\dot{x}-i\dot{y}%
)^{2}+i\left( 1-\frac{z}{\sqrt{w}}\right) F_{1}(z+\sqrt{w})+ \\
&&+i\left( -1-\frac{z}{\sqrt{w}}\right) F_{2}(z-\sqrt{w}).
\end{eqnarray*}%
\bigskip

6) $\alpha=1$, $\beta =\mp i$ and $\gamma =A=B=C=0$. Then
\begin{equation*}
C_{ab}=\left(
\begin{array}{cc}
2y & -x\pm iy \\
-x\pm iy & \mp 2ix%
\end{array}%
\right)
\end{equation*}%
and equation (\ref{eq.Hie2}) becomes
\begin{equation}
(x\mp iy)\left( V_{,xx}-V_{,yy}\right) +2\left( y\pm ix\right) V_{,xy}\pm
3iV_{,y}+3V_{,x}=0  \label{eq.Hie6a}
\end{equation}%
from which follows%
\begin{equation}
V_{26}=\frac{F_{1}(z)}{r}+F_{2}^{\prime }(z)  \label{eq.Hie6b}
\end{equation}%
where $F_{2}^{\prime }=\frac{dF_{2}}{dz}$ and $z=x\pm iy$.

We find the QFI (see (3.2.18) of \cite{Hietarinta 1987})
\begin{equation}
I_{61}=(y\dot{x}-x\dot{y})(\dot{x}\pm i\dot{y})-izV+iF_{2}(z).
\label{eq.Hie6c}
\end{equation}%
\bigskip

7) $AB\neq 0$, $A\neq B$ and $\alpha=\beta =\gamma =C=0$. Then $C_{ab}=\left(
\begin{array}{cc}
A & 0 \\
0 & B%
\end{array}%
\right) $.

Equation (\ref{eq.Hie2}) becomes
\begin{equation}  \label{eq.Hie7a}
(A-B)V_{,xy}=0 \implies V_{,xy}=0
\end{equation}
which gives the separable potential
\begin{equation}  \label{eq.Hie7b}
V_{27}= F_{1}(x) + F_{2}(y).
\end{equation}

We find the irreducible QFIs (see (3.2.20) of \cite{Hietarinta 1987})
\begin{equation*}
I_{71a}=\dot{x}^{2}+2F_{1}(x), \enskip I_{71b}=\dot{y}^{2}+2F_{2}(y).
\end{equation*}

It can be shown that there are four special potentials of the potential (\ref%
{eq.Hie7b}) which admit additional time-dependent QFIs and are
superintegrable. These are:

7a. The potential
\begin{equation*}
V_{271} =\frac{k_{1}}{\left( x+c_{1}\right) ^{2}}+\frac{k_{2}}{\left(
y+c_{2}\right) ^{2}}
\end{equation*}%
admits the independent FIs
\begin{eqnarray*}
I_{72a} &=& -\frac{t^{2}}{2} \dot{y}^{2} + t (y+c_{2})\dot{y} - t^{2}\frac{%
k_{2}}{(y+ c_{2})^{2}} - \frac{1}{2}y^{2} - c_{2}y \\
I_{72b} &=& -\frac{t^{2}}{2} \dot{x}^{2} + t (x+c_{1})\dot{x} - t^{2}\frac{%
k_{1}}{(x+ c_{1})^{2}} - \frac{1}{2}x^{2} - c_{1}x.
\end{eqnarray*}

7b. The potential
\begin{equation*}
V_{272}=F_{1}(x)+\frac{k_{2}}{\left( y+c_{2}\right) ^{2}}
\end{equation*}%
admits the FI $I_{72a}$.

7c. The potential
\begin{equation*}
V_{273}=F_{2}(y)+ \frac{k_{1}}{\left( x+c_{1}\right) ^{2}}
\end{equation*}%
admits the FI $I_{72b}$.

7d. The potential (see \cite{Fris})
\begin{equation*}
V_{274}=-\frac{\lambda ^{2}}{8}(x^{2}+y^{2})-\frac{\lambda ^{2}}{4}\left(
c_{1}x+ c_{2}y\right) - \frac{k_{1}}{(x+c_{1})^{2}}-\frac{k_{2}}{%
(y+c_{2})^{2}}
\end{equation*}
admits the independent FIs
\begin{eqnarray*}
I_{73a} &=& e^{\lambda t}\left[ -\dot{x}^{2}+\lambda (x+c_{1})\dot{x}-\frac{%
\lambda^{2}}{4}(x+c_{1})^{2}+ \frac{2k_{1}}{(x+c_{1})^{2}}\right] \\
I_{73b} &=& e^{\lambda t}\left[ -\dot{y}^{2}+\lambda (y+c_{2})\dot{y}-\frac{%
\lambda^{2}}{4}(y+c_{2})^{2}+\frac{2k_{2}}{(y+c_{2})^{2}}\right].
\end{eqnarray*}

In all the above relations $\lambda, c_{1}, c_{2}, k_{1}, k_{2}$ are arbitrary constants.
\bigskip

8) $C\neq0$ and $\alpha=\beta=\gamma=0$.

Then $C_{ab}=\left(
\begin{array}{cc}
A & C \\
C & B%
\end{array}%
\right) $ and equation (\ref{eq.Hie2}) becomes
\begin{equation}
C(V_{,yy}-V_{,xx})+(A-B)V_{,xy}=0.  \label{eq.Hie8a}
\end{equation}

Solving (\ref{eq.Hie8a}) we find the potential
\begin{equation}  \label{eq.Hie8b}
V_{28} = F_{1}\left(y + b_{0}x + \sqrt{b_{0}^{2}+1}x\right) + F_{2}\left(y +
b_{0}x - \sqrt{b_{0}^{2}+1}x\right)
\end{equation}
where $b_{0} \equiv \frac{A-B}{2C}$.

This potential admits the QFI
\begin{equation}  \label{eq.Hie8bb}
I_{81}= A\dot{x}^{2} + B\dot{y}^{2} + 2C\dot{x}\dot{y} + (A+B)V + 2C \sqrt{%
b_{0}^{2}+1} (F_{1}- F_{2}).
\end{equation}
We note that $b_{0}(A,B,C)$. Here $A,B,C$ are parameters of the potential
and therefore cannot be taken as independent parameters of the QFI.

For $b_{0}=0$ we have $A=B$, $V_{,yy} - V_{,xx}=0$ and the potential reduces
to
\begin{equation}  \label{eq.Hie8c}
V_{28}(b_{0}=0)= F_{1}(y+x) + F_{2}(y-x)
\end{equation}
which is the solution of the 1d-wave equation.

For the potential (\ref{eq.Hie8c}) we find the QFI
\begin{equation*}
I_{82}= \dot{x}\dot{y} + F_{1}(y+x) - F_{2}(y-x).
\end{equation*}
\bigskip

9) $A=2$, $C=\pm i$ and $\alpha=\beta =\gamma =B=0$. Then
\begin{equation*}
C_{ab}=\left(
\begin{array}{cc}
2 & \pm i \\
\pm i & 0%
\end{array}%
\right)
\end{equation*}%
and equation (\ref{eq.Hie2}) becomes
\begin{equation}
\mp i(V_{,xx}-V_{,yy})+2V_{,xy}=0.  \label{eq.Hie9a}
\end{equation}

Solving (\ref{eq.Hie9a}) we find the potential
\begin{equation}  \label{eq.Hie9b}
V_{29} = r^{2} F_{1}^{\prime \prime }(z) + F_{2}(z)
\end{equation}
where $F_{1}^{\prime \prime }= \frac{d^{2}F_{1}}{dz^{2}}$ and $z=x\pm iy$.

This potential admits the QFI (see (3.2.21) of \cite{Hietarinta 1987})
\begin{equation}
I_{91}=\dot{x}(\dot{x}\pm i\dot{y})+V_{29}+ 2zF_{1}^{\prime}(z) -2F_{1}(z).
\label{eq.Hie9c}
\end{equation}

Observe that for the trivial KT $C_{ab}=A\delta _{ab}$ the condition $%
G_{,a}=2C_{ab}V^{,b}$ gives
\begin{equation*}
G_{,a}=2AV_{,a}\implies G=2AV
\end{equation*}%
for all potentials $V(x,y)$. Therefore we recover the trivial result that
all 2d-potentials $V(x,y)$ admit the QFI
\begin{equation*}
I=A(\dot{x}^{2}+\dot{y}^{2}+2V)=2AH.
\end{equation*}
\bigskip

Comparing with previous works we see that the potentials $V_{21a}$ and $%
V_{28}$ are new. The potential $V_{274}(c_{1}=c_{2}=0)$ is mentioned in \cite%
{Fris}.

\subsection{The superintegrable potentials}

\label{sec.super}

When a potential belongs to two of the above 9 \textbf{Class II} cases
simultaneously is superintegrable (e.g. potentials (3.2.34)- (3.2.36) of
\cite{Hietarinta 1987}) because in that case the potential admits two more autonomous FIs other than the Hamiltonian. From the above results we
find the following \textbf{Class II} superintegrable potentials
(see also \cite{Ranada 1997}, \cite{Kalnins 2001}). \bigskip

S1) The potential (see (3.2.34) in \cite{Hietarinta 1987}, case (b) in \cite%
{Ranada 1997} and \cite{Fris})
\begin{equation}
V_{s1}=\frac{k}{2}(x^{2}+y^{2})+\frac{b}{x^{2}}+\frac{c}{y^{2}}
\label{eq.Hie10a}
\end{equation}
where $k,b,c$ are arbitrary constants.

This is of the form (\ref{eq.Hie3b}) for $d_{1}=d_{2}=1$,
\begin{equation*}
F_{1}\left( \frac{y}{x}\right) =b\left( \frac{y}{x}\right) ^{2}+c\left(
\frac{x}{y}\right) ^{2},\enskip F_{2}(x^{2}+y^{2})=\frac{k}{2}(x^{2}+y^{2})+%
\frac{b+c}{x^{2}+y^{2}}
\end{equation*}%
and also of the separable form (\ref{eq.Hie7b}). Therefore $V_{s1}$ admits
the additional QFIs
\begin{eqnarray}
I_{s1a} &=&(y\dot{x}-x\dot{y})^{2}+2b\frac{y^{2}}{x^{2}}+2c\frac{x^{2}}{y^{2}%
}  \label{eq.Hie10b} \\
I_{s1b} &=&\frac{1}{2}\dot{x}^{2}+\frac{k}{2}x^{2}+\frac{b}{x^{2}}
\label{eq.Hie10c} \\
I_{s1c} &=&\frac{1}{2}\dot{y}^{2}+\frac{k}{2}y^{2}+\frac{c}{y^{2}}.
\label{eq.Hie10d}
\end{eqnarray}%
We note that $V_{s1}\left( k=-\frac{\lambda^{2}}{4}, b=-k_{1}, c=-k_{2}
\right)$, $\lambda \neq 0$, is the $V_{274}$ for $c_{1}=c_{2}=0$ and
therefore admits also the time-dependent FIs $I_{73a}$, $I_{73b}$. \bigskip

S2) Potentials of the form (\ref{eq.Hie4b}) and (\ref{eq.Hie7b}).
 Then we have to solve the systems of PDEs (\ref{eq.Hie4a})
and $V_{,xy}=0$. We find
\begin{equation}
V_{s2}=\frac{k_{1}}{2}(x^{2}+4y^{2})+\frac{k_{2}}{x^{2}}+k_{3}y
\label{eq.Hie11a}
\end{equation}%
and the QFIs
\begin{eqnarray}
I_{s2a} &=&\dot{x}(y\dot{x}-x\dot{y})-k_{1}yx^{2}+\frac{2k_{2}y}{x^{2}}-%
\frac{k_{3}}{2}x^{2}  \label{eq.Hie11b} \\
I_{s2b} &=&\frac{1}{2}\dot{x}^{2}+\frac{k_{1}}{2}x^{2}+\frac{k_{2}}{x^{2}}
\label{eq.Hie11c} \\
I_{s2c} &=&\frac{1}{2}\dot{y}^{2}+2k_{1}y^{2}+k_{3}y.  \label{eq.Hie11d}
\end{eqnarray}
where $k_{1}, k_{2}, k_{3}$ are arbitrary constants.

This is the superintegrable potential of case (a) of \cite{Ranada 1997}.
Note that the QFI\ $I_{3}^{a}$ given in \cite{Ranada 1997} is not correct.
The correct is the $I_{s2a}$ of (\ref{eq.Hie11b}) above.

\bigskip We remark that the potential (3.2.35) in \cite{Hietarinta 1987} is
superintegrable only for $b=4a$ in which case the potential becomes $V_{s2}$
for $k_{1}=2a$, $k_{2}=c$ and $k_{3}=0$. \bigskip

S3) Potentials of the form (\ref{eq.Hie3b}) and (\ref{eq.Hie4b}). We solve
 the system of PDEs (\ref{eq.Hie3a}) and (\ref{eq.Hie4a}). We
find
\begin{equation}
V_{s3} = \frac{k_{1}}{x^{2}} + \frac{k_{2}}{r} + \frac{k_{3}y}{rx^{2}}
\label{eq.Hie12a}
\end{equation}
and the QFIs
\begin{eqnarray}
I_{s3a} &=& (y\dot{x} - x\dot{y})^{2} + 2k_{1}\frac{y^{2}}{x^{2}} + 2k_{3}%
\frac{ry}{x^{2}}  \label{eq.Hie12b} \\
I_{s3b} &=& \dot{x}(y\dot{x} - x\dot{y}) + 2k_{1} \frac{y}{x^{2}} + k_{2}%
\frac{y}{r} + k_{3}\frac{x^{2}+2y^{2}}{rx^{2}}  \label{eq.Hie12c}
\end{eqnarray}
where $r^{2}=x^{2}+y^{2}$.

The superintegrable potential (\ref{eq.Hie12a}) is symmetric ($x
\leftrightarrow y$) to the superintegrable potential of case (c) of \cite%
{Ranada 1997}. Indeed in order to find the superintegrable potential of \cite%
{Ranada 1997} we simply consider the case leading to the potential of the
form $V_{24}$ of (\ref{eq.Hie4b}) for $\beta=1$ instead of $\alpha=1$.

We note that if we rename the constants in (\ref{eq.Hie12a}) as $k_{1}=b+c$,
$k_{2}=a$, $k_{3}=c-b$ we recover the superintegrable potential (3.2.36) of
{\large \cite{Hietarinta 1987}}. Indeed we have
\begin{equation*}
V_{s3}=\frac{a}{r}+\frac{\frac{b}{r+y}+\frac{c}{r-y}}{r}.
\end{equation*}

S4) If we substitute the solution (\ref{eq.Hie4b}) of the PDE (\ref{eq.Hie4a}%
) in the PDE (\ref{eq.Hie4a.1}), we find that for
\begin{equation*}
F_{1}(r+y)=k_{1}+k_{2}\sqrt{r+y}, \enskip F_{2}(r-y)=k_{3}\sqrt{r-y}
\end{equation*}
both PDEs (\ref{eq.Hie4a}) and (\ref{eq.Hie4a.1}) are satisfied
simultaneously. Therefore the potential (see case (d) in \cite{Ranada 1997})
\begin{equation}
V_{s4} = \frac{k_{1}}{r} + k_{2} \frac{\sqrt{r+y}}{r} + k_{3}\frac{\sqrt{r-y}%
}{r}  \label{eq.Hie12d}
\end{equation}
is superintegrable with additional QFIs
\begin{eqnarray*}
I_{s4a} &=& \dot{x}(y\dot{x}-x\dot{y}) + \frac{k_{1}y}{r} + \frac{k_{3}(r+y)%
\sqrt{r-y}-k_{2}(r-y)\sqrt{r+y}}{r}  \label{eq.Hie12e} \\
I_{s4b} &=& \dot{y}(x\dot{y}-y\dot{x}) + G(x,y)  \label{eq.Hie12f}
\end{eqnarray*}
where $G_{,x} + yV_{s4,y}=0$ and $G_{,y} + yV_{s4,x} -2xV_{s4,y}=0$.

We note that in the case (d) in \cite{Ranada 1997} the corresponding QFIs $%
I_{2}^{d}$ and $I_{3}^{d}$ are not correct, because $\{H,I_{2}^{d}\}\neq0$
and $\{H,I_{3}^{d}\}\neq0$. Moreover this superintegrable potential is the
case (E20) in \cite{Kalnins 2001}; and it is not mentioned in the review
\cite{Hietarinta 1987}.

In the following tables we collect the results on \textbf{Class II}
potentials \ with the corresponding reference to the review paper
\cite{Hietarinta 1987}. \bigskip

\begin{tabular}{|l|l|l|}
\hline
\multicolumn{3}{|c|}{Integrable potentials} \\ \hline
{\large Potential} & {\large Ref \cite{Hietarinta 1987}} & {\large LFIs and
QFIs} \\ \hline
$V_{21}=\frac{F_{1}\left( \frac{y}{x}\right) }{x^{2}+y^{2}}%
+F_{2}(x^{2}+y^{2})$ & (3.2.10) & $I_{11}=(y\dot{x}-x\dot{y}%
)^{2}+2F_{1}\left( \frac{y}{x}\right) $ \\ \hline
$V_{21a}= \frac{k}{x^{2}+\ell y^{2}} + F_{2}(x^{2}+y^{2})$ & - & $I_{11a} =
(y\dot{x}-x\dot{y})^{2} + \frac{2k(1-\ell) y^{2}}{x^{2}+\ell y^{2}}$ \\
\hline
\makecell[l]{$V_{22}=\frac{F_{1}(u)-F_{2}(v)}{u^{2}-v^{2}}$, \\
$u^{2}=r^{2}+A+\left[ (r^{2}+A)^{2}-4Ax^{2}\right] ^{1/2}$, \\
$v^{2}=r^{2}+A-\left[ (r^{2}+A)^{2}-4Ax^{2}\right] ^{1/2}$} & (3.2.7,8) & $%
I_{21}=(y\dot{x}-x\dot{y})^{2}+A\dot{x}^{2}+\frac{v^{2}F_{1}(u)-u^{2}F_{2}(v)%
}{u^{2}-v^{2}}$ \\ \hline
\makecell[l]{$V_{23}=\frac{F_{1}(u)-F_{2}(v)}{u^{2}-v^{2}}$, \\
$u^{2}=r^{2}+\left[ r^{4}-4A(x\pm iy)^{2}\right] ^{1/2}$, \\
$v^{2}=r^{2}-\left[ r^{4}-4A(x\pm iy)^{2}\right] ^{1/2}$} & (3.2.7,12) & $%
I_{31}=(y\dot{x}-x\dot{y})^{2}+A(\dot{x}\pm i\dot{y})^{2}+\frac{%
v^{2}F_{1}(u)-u^{2}F_{2}(v)}{u^{2}-v^{2}}$ \\ \hline
$V_{24}=\frac{F_{1}(r+y)+F_{2}(r-y)}{r}$ & (3.2.15) & $I_{41}=\dot{x}(y\dot{x%
}-x\dot{y})+\frac{(r+y)F_{2}(r-y)-(r-y)F_{1}(r+y)}{r}$ \\ \hline
\makecell[l]{$V_{25}=w^{-1/2}\left[
F_{1}(z+\sqrt{w})+F_{2}(z-\sqrt{w})\right] $, \\ $z=x+iy$, $w=x-iy$} &
(3.2.17) & \makecell[l]{$I_{51}=(y\dot{x}-x\dot{y})(\dot{x}+i\dot{y})+
\frac{i}{8}(\dot{x}-i\dot{y})^{2}+$ \\ \qquad \enskip $+i\left(
1-\frac{z}{\sqrt{w}}\right) F_{1}(z+\sqrt{w})+$ \\ \qquad \enskip $+i\left(
-1-\frac{z}{\sqrt{w}}\right) F_{2}(z-\sqrt{w})$} \\ \hline
\makecell[l]{$V_{26}=\frac{F_{1}(z)}{r}+F_{2}^{\prime }(z)$, \\
$F_{2}^{\prime }=\frac{dF_{2}}{dz}$, $z=x\pm iy$} & (3.2.18) & $I_{61}=(y%
\dot{x}-x\dot{y})(\dot{x}\pm i\dot{y})-izV+iF_{2}(z)$ \\ \hline
$V_{27}=F_{1}(x)+F_{2}(y)$ & (3.2.20) & $I_{71a}=\frac{1}{2}\dot{x}%
^{2}+F_{1}(x)$, $I_{71b}=\frac{1}{2}\dot{y}^{2}+F_{2}(y)$ \\ \hline
\makecell[l]{$V_{28}=F_{1}\left( y+b_{0}x+\sqrt{b_{0}^{2}+1}x\right) +$ \\
\qquad \enskip $+ F_{2}\left(y+b_{0}x-\sqrt{b_{0}^{2}+1}x\right) $,
$b_{0}\equiv \frac{A-B}{2C}$} & - & \makecell[l]{$I_{81}=A\dot{x}^{2}+B%
\dot{y}^{2}+$ \\ \qquad \enskip $+ 2C\dot{x}\dot{y}+(A+B)V+
2C\sqrt{b_{0}^{2}+1}(F_{1}-F_{2})$} \\ \hline
$V_{28}(b_{0}=0)=F_{1}(y+x)+F_{2}(y-x)$ & - & $I_{82}=\dot{x}\dot{y}%
+F_{1}(y+x)-F_{2}(y-x)$ \\ \hline
\makecell[l]{$V_{29}=r^{2}F_{1}^{\prime \prime }(z)+F_{2}(z)$,
$F_{1}^{\prime \prime }=\frac{d^{2}F_{1}}{dz^{2}}$, \\ $z=x\pm iy$} &
(3.2.21) & $I_{91}=\dot{x}(\dot{x}\pm i\dot{y})+V_{29} +2zF_{1}^{\prime
}(z)- 2F_{1}(z)$ \\ \hline
\end{tabular}

\bigskip

\begin{tabular}{|l|l|l|}
\hline
\multicolumn{3}{|c|}{Superintegrable potentials} \\ \hline
{\large Potential} & {\large Ref \cite{Hietarinta 1987}} & {\large LFIs and
QFIs} \\ \hline
$V_{s1}= \frac{k}{2}(x^{2}+y^{2}) + \frac{b}{x^{2}} + \frac{c}{y^{2}}$ &
(3.2.34) & \makecell[l]{$I_{s1a}= (y\dot{x}-x\dot{y})^{2} + 2b
\frac{y^{2}}{x^{2}} + 2c \frac{x^{2}}{y^{2}}$, \\ $I_{s1b} =
\frac{1}{2}\dot{x}^{2} + \frac{k}{2}x^{2} + \frac{b}{x^{2}}$, $I_{s1c} =
\frac{1}{2}\dot{y}^{2} + \frac{k}{2}y^{2} + \frac{c}{y^{2}}$ \\ - For $k=0$:
$I_{72a}$, $I_{72b}$ \\ where $c_{1}=c_{2}=0, k_{1}=b, k_{2}=c$ \\ - For
$k=-\frac{\lambda^{2}}{4}\neq0$: $I_{73a}$, $I_{73b}$ \\ where
$c_{1}=c_{2}=0, k_{1}=-b, k_{2}=-c$ } \\ \hline
$V_{s2}= \frac{k_{1}}{2}(x^{2}+4y^{2}) + \frac{k_{2}}{x^{2}} + k_{3}y$ & %
\makecell[l]{(3.2.35) \\ for $k_{3}=0$} & \makecell[l]{$I_{s2a} =
\dot{x}(y\dot{x}-x\dot{y}) - k_{1}yx^{2} + \frac{2k_{2}y}{x^{2}} -
\frac{k_{3}}{2} x^{2}$, \\ $I_{s2b}= \frac{1}{2}\dot{x}^{2} +
\frac{k_{1}}{2}x^{2} + \frac{k_{2}}{x^{2}}$, $I_{s2c}=
\frac{1}{2}\dot{y}^{2} + 2k_{1}y^{2}+ k_{3}y$} \\ \hline
$V_{s3} = \frac{k_{1}}{x^{2}} + \frac{k_{2}}{r} + \frac{k_{3}y}{rx^{2}}$ &
(3.2.36) & \makecell[l]{$I_{s3a}= (y\dot{x} - x\dot{y})^{2} +
2k_{1}\frac{y^{2}}{x^{2}} + 2k_{3}\frac{ry}{x^{2}}$, \\ $I_{s3b}=
\dot{x}(y\dot{x} - x\dot{y}) + 2k_{1} \frac{y}{x^{2}} + k_{2}\frac{y}{r} +
k_{3}\frac{x^{2}+2y^{2}}{rx^{2}}$} \\ \hline
$V_{s4} = \frac{k_{1}}{r} + k_{2} \frac{\sqrt{r+y}}{r} + k_{3}\frac{\sqrt{r-y%
}}{r}$ & - & \makecell[l]{$I_{s4a} = \dot{x}(y\dot{x}-x\dot{y}) +
\frac{k_{1}y}{r} + \frac{k_{3}(r+y)\sqrt{r-y}-k_{2}(r-y)\sqrt{r+y}}{r}$, \\
$I_{s4b}= \dot{y}(x\dot{y}-y\dot{x}) + G(x,y)$} \\ \hline
$V_{271}=\frac{k_{1}}{\left( x+c_{1} \right)^{2}}+ \frac{k_{2}}{\left(
y+c_{2} \right)^{2}}$ & (3.2.20) & \makecell[l]{$I_{71a}$, $I_{71b}$, \\
$I_{72a}=-\frac{t^{2}}{2}\dot{y}^{2}+t(y+c_{2})\dot{y}-
t^{2}\frac{k_{2}}{(y+c_{2})^{2}}-\frac{1}{2}y^{2}-c_{2}y$, \\
$I_{72b}=-\frac{t^{2}}{2}\dot{x}^{2}+t(x+c_{1})\dot{x}-
t^{2}\frac{k_{1}}{(x+c_{1})^{2}}-\frac{1}{2}x^{2}-c_{1}x$} \\ \hline
$V_{272}=F_{1}(x)+\frac{k_{2}}{\left( y+c_{2}\right) ^{2}}$ & (3.2.20) & $%
I_{71a}$, $I_{71b}$, $I_{72a}$ \\ \hline
$V_{273}=F_{2}(y)+\frac{k_{1}}{\left( x+c_{1}\right) ^{2}}$ & (3.2.20) & $%
I_{71a}$, $I_{71b}$, $I_{72b}$ \\ \hline
\makecell[l]{$V_{274}=-\frac{\lambda ^{2}}{8}(x^{2}+y^{2})-\frac{\lambda
^{2}}{4}\left( c_{1}x+ c_{2}y\right)-$ \\ \qquad \quad
$-\frac{k_{1}}{(x+c_{1})^{2}}-\frac{k_{2}}{(y+c_{2})^{2}}$, $\lambda \neq0$}
& (3.2.20) & \makecell[l]{$I_{71a}$, $I_{71b}$, \\ $I_{73a}=e^{\lambda
t}\left[ -\dot{x}^{2}+\lambda (x+c_{1})\dot{x}-\frac{\lambda
^{2}}{4}(x+c_{1})^{2}+\frac{2k_{1}}{(x+c_{1})^{2}}\right] $, \\
$I_{73b}=e^{\lambda t}\left[ -\dot{y}^{2}+\lambda
(y+c_{2})\dot{y}-\frac{\lambda
^{2}}{4}(y+c_{2})^{2}+\frac{2k_{2}}{(y+c_{2})^{2}}\right]$} \\ \hline
\end{tabular}

\section{The constraint $\left(L_{b}V^{,b}\right)_{,a} = -2 L_{(a;b)} V^{,b}
- \protect\lambda^{2}L_{a}$}

\label{sec.const3}

The integrability condition of the constraint $\left(
L_{b}V^{,b}\right)_{,a}=-2L_{(a;b)}V^{,b}-\lambda ^{2}L_{a}$ gives the PDE (%
\ref{eq.PDE3.3}).

As mentioned above in section \ref{sec.find.Pots} in order to find new
potentials from the PDE (\ref{eq.PDE3.3}) one (or both) of the conditions $\alpha= \beta =0$ and $a_{1}=a_{3}$ must be relaxed. However, if we do find a new
potential, this solution should satisfy also the remaining PDEs (\ref%
{eq.PDE3.1}) and (\ref{eq.PDE3.2}) in order to admit the time-dependent QFI $%
I_{3}$ given in case \textbf{Integral 3} of theorem \ref{The first integrals
of an autonomous holonomic dynamical system}. New potentials which admit the
QFI $I_{3}$ shall be referred as \textbf{Class III} potentials.

We note that the PB $\{H,I_{3}\}= \frac{\partial I_{3}}{\partial t} \neq 0$.
Therefore to find a new integrable potential we should find a \textbf{Class
III} potential admitting two independent FIs of the form $I_{3}$, say $%
I_{3a} $ and $I_{3b}$, such that $\{I_{3a}, I_{3b}\} =0$.

After relaxing one, or both, of the conditions $\alpha=\beta =0$ and $a_{1}=a_{3}$ we found that the only non-trivial \textbf{Class III}
potential is the superintegrable potential $V_{3b}=-\frac{\lambda ^{2}}{2}%
(x^{2}+y^{2})$ (see subsection \ref{subsec.V3}) found for $\alpha\neq 0$ or $\beta\neq 0$ above. Therefore there are no new \textbf{Class III} potentials.

\section{Using FIs to find the solution of 2d integrable dynamical systems}

 In this section we consider examples which show how one uses the 2d (super-)integrable potentials to find the solution of the dynamical equations.
\bigskip

1) The superintegrable potential $V_{3b}=-\frac{1}{2}k^{2}(x^{2}+y^{2})$ where $k\neq 0$.\\

 We find the solution by using the time-dependent LFIs $L_{42\pm}= e^{\pm kt}(\dot{x}\mp kx)$ and $L_{43\pm}=e^{\pm kt}(\dot{y} \mp ky)$. Specifically we have
\[
\begin{cases}
e^{kt}(\dot{x} -kx) = c_{1+} \\
e^{-kt}(\dot{x} +kx) = c_{1-}
\end{cases}
\implies
\begin{cases}
\dot{x} -kx = c_{1+}e^{-kt} \\
\dot{x} +kx = c_{1-}e^{kt}
\end{cases}
\implies
x(t) = \frac{c_{1-}}{2k}e^{kt} - \frac{c_{1+}}{2k}e^{-kt}.
\]
Similarly for the LFIs $L_{43\pm}$ we find
\[
y(t) = \frac{c_{2-}}{2k}e^{kt} - \frac{c_{2+}}{2k}e^{-kt}.
\]
Here $c_{1\pm}, c_{2\pm}$ are arbitrary constants
\bigskip

2) The integrable potential $V_{2}=cy+F(x)$ where $F''\neq0$.\\

Using the LFI $L_{31}=\dot{y}+ct=c_{1}$ we find directly $y(t)= -\frac{c}{2}t^{2} +c_{1}t + c_{2}$ where $c,c_{1},c_{2}$ are arbitrary constants.

Using the QFI $2Q_{31} =\dot{x}^{2}+2F(x)=const=c_{3}$ we have
\[
\frac{dx}{dt} = \pm \left[-2F(x)+c_{3}\right]^{1/2} \implies dt= \pm \left[-2F(x)+c_{3}\right]^{-1/2}dx \implies t= \pm \int\left[-2F(x)+c_{3}\right]^{-1/2}dx + c_{0}
\]
where $c_{0}$ is an arbitrary constant. The inverse function of $t=t(x)$ is the solution of the system. If the function $F(x)$ is given, the solution can be explicitly determined.
\bigskip

3) For the integrable potential $V_{27}= F_{1}(x) + F_{2}(y)$ by using the QFIs
\[
I_{71a}= \frac{1}{2}\dot{x}^{2} + F_{1}(x) \enskip \text{and} \enskip I_{71b}=\frac{1}{2}\dot{y}^{2} + F_{2}(y)
\]
we find
\[
t= \int\left[c_{1}-2F_{1}(x)\right]^{-1/2}dx + c_{0}, \enskip t= \int\left[c_{2}-2F_{2}(y)\right]^{-1/2}dy + c_{3}
\]
where $c_{0}, c_{1}= 2I_{71a}, c_{2}=2I_{71b}, c_{3}$ are constants.

\section{Conclusions}

\label{sec.conclusions}

Using Theorem \ref{The first integrals of an autonomous holonomic dynamical
system} we have reproduced in a systematic way most known integrable
and superintegrable 2d potentials of autonomous conservative dynamical
systems. The method used being covariant it is directly applicable to spaces
of higher dimensions and to metrics with any signature and curvature.

We have found two classes of potentials and in each class we have determined
the integrable and the superintegrable potentials together with their QFIs. Since
the general solution of the PDE (\ref{eq.PDE2}) is not possible we have
found the potentials due to certain solutions only. New solutions of this
equation will lead to new integrable and possibly superintegrable 2d
potentials.

It appears that the most difficult part in the application of Theorem \ref%
{The first integrals of an autonomous holonomic dynamical system} to higher
dimensions and curved configuration spaces is the determination of the KTs. The use
of algebraic computing is limited once one considers higher dimensions since
then the number of the components of the KT increases dramatically.
Fortunately, today new techniques in Differential Geometry have been
developed (e.g. \cite{Rani 2003}, \cite{Kalnins 1980}, \cite{KTs and CKVs 2006}, \cite{Crampin 2008}, \cite{Glass 2010}),
 especially in
the case of spaces of constant curvature and decomposable spaces, which can
help to deal with this problem.

\bigskip
{ {\textbf{Acknowledgements:}}}
A.P. acknowledges financial support of Agencia Nacional de Investigaci\'{o}n
y Desarrollo - ANID through the program FONDECYT Iniciaci\'{o}n grant no.
11180126. Additionally, by Vicerrector\'{\i}a de Investigaci\'{o}n y
Desarrollo Tecnol\'{o}gico at Universidad Catolica del Norte.

\end{document}